\documentstyle[12pt]{article}
\catcode`\@=11
\long\def\@makefntext#1{
\protect\noindent \hbox to 3.2pt {\hskip-.9pt
$^{{\ninerm\@thefnmark}}$\hfil}#1\hfill}              %CAN BE USED

\def\@makefnmark{\hbox to 0pt{$^{\@thefnmark}$\hss}}  %ORIGINAL

\def\ps@myheadings{\let\@mkboth\@gobbletwo
\def\@oddhead{\hbox{}
\rightmark\hfil\ninerm\thepage}
\def\@oddfoot{}\def\@evenhead{\ninerm\thepage\hfil
\leftmark\hbox{}}\def\@evenfoot{}
\def\sectionmark##1{}\def\subsectionmark##1{}}

\renewcommand{\thefootnote}{\fnsymbol{footnote}}

\newcounter{sectionc}
\newcounter{subsectionc}
\newcounter{subsubsectionc}
\renewcommand{\section}[1] {\vspace*{0.6cm}\addtocounter{sectionc}{1}
\setcounter{subsectionc}{0}\setcounter{subsubsectionc}{0}\noindent
        {\normalsize\bf\thesectionc. #1}\par\vspace*{0.4cm}}
\renewcommand{\subsection}[1] {\vspace*{0.6cm}
        \addtocounter{subsectionc}{1}
        \setcounter{subsubsectionc}{0}\noindent
        {\normalsize\it\thesectionc.\thesubsectionc. #1}
        \par\vspace*{0.4cm}}
\renewcommand{\subsubsection}[1]
{\vspace*{0.6cm}\addtocounter{subsubsectionc}{1}
        \noindent
{\normalsize\rm\thesectionc.\thesubsectionc.\thesubsubsectionc.
        #1}\par\vspace*{0.4cm}}

\newcounter{appendixc}
\newcounter{subappendixc}[appendixc]
\newcounter{subsubappendixc}[subappendixc]

\renewcommand{\appendix}[1] {\vspace*{0.6cm}
        \refstepcounter{appendixc}
        \setcounter{figure}{0}
        \setcounter{table}{0}
        \setcounter{equation}{0}
        \renewcommand{\thefigure}{\Alph{appendixc}.\arabic{figure}}
        \renewcommand{\thetable}{\Alph{appendixc}.\arabic{table}}
        \renewcommand{\theappendixc}{\Alph{appendixc}}
        \renewcommand{\theequation}
        {\Alph{appendixc}.\arabic{equation}}
        \noindent{\bf Appendix \theappendixc #1}\par\vspace*{0.4cm}}

\def\abstracts#1{{

\centering{\begin{minipage}{12.2truecm}
        \footnotesize\baselineskip=12pt\noindent
        \centerline{\footnotesize ABSTRACT}\vspace*{0.3cm}
        \parindent=0pt #1
        \end{minipage}}\par}}

\renewenvironment{thebibliography}[1]
        {\begin{list}{\arabic{enumi}.}
        {\usecounter{enumi}\setlength{\parsep}{0pt}
\setlength{\leftmargin 1.25cm}{\rightmargin 0pt}
         \setlength{\itemsep}{0pt} \settowidth
        {\labelwidth}{#1.}\sloppy}}{\end{list}}

\newcounter{itemlistc}
\newcounter{romanlistc}
\newcounter{alphlistc}
\newcounter{arabiclistc}

\newcommand{\fcaption}[1]{
        \refstepcounter{figure}
        \setbox\@tempboxa = \hbox{\footnotesize Fig.~\thefigure. #1}
        \ifdim \wd\@tempboxa > 6in
           {\begin{center}
        \parbox{6in}{\footnotesize
        \baselineskip=12pt Fig.~\thefigure. #1}
            \end{center}}
        \else
             {\begin{center}
             {\footnotesize Fig.~\thefigure. #1}
              \end{center}}
        \fi}

\newcommand{\tcaption}[1]{
        \refstepcounter{table}
        \setbox\@tempboxa = \hbox{\footnotesize Table~\thetable. #1}
        \ifdim \wd\@tempboxa > 6in
           {\begin{center}
        \parbox{6in}
        {\footnotesize\baselineskip=12pt Table~\thetable. #1}
            \end{center}}
        \else
             {\begin{center}
             {\footnotesize Table~\thetable. #1}
              \end{center}}
        \fi}

\def\@citex[#1]#2{\if@filesw\immediate\write\@auxout
        {\string\citation{#2}}\fi
\def\@citea{}\@cite{\@for\@citeb:=#2\do
        {\@citea\def\@citea{,}\@ifundefined
        {b@\@citeb}{{\bf ?}\@warning
        {Citation `\@citeb' on page \thepage \space undefined}}
        {\csname b@\@citeb\endcsname}}}{#1}}

\newif\if@cghi
\def\cite{\@cghitrue\@ifnextchar [{\@tempswatrue
        \@citex}{\@tempswafalse\@citex[]}}
\def\citelow{\@cghifalse\@ifnextchar [{\@tempswatrue
        \@citex}{\@tempswafalse\@citex[]}}
\def\@cite#1#2{{$\null^{#1}$\if@tempswa\typeout
        {IJCGA warning: optional citation argument
        ignored: `#2'} \fi}}

 1
 1
 1

\font\ninerm=cmr9

\input epsf
\global\arraycolsep=2pt

\begin{document}

\begin{titlepage}

\begin{flushright}
CERN-TH/95-307\\
hep-ph/9511409
\end{flushright}

\vspace{3.0cm}

\begin{center}
\Large\bf Heavy Flavour Physics
\end{center}

\vspace{1.0cm}

\begin{center}
Matthias Neubert\\
{\sl Theory Division, CERN, CH-1211 Geneva 23, Switzerland}
\end{center}

\vspace{2.0cm}

\begin{abstract}
The current status of the theory and phenomenology of weak decays of
hadrons containing a heavy quark is reviewed. Exclusive semileptonic
and rare decays of $B$ mesons are discussed, as well as inclusive
decay rates, the semileptonic branching ratio of the $B$ meson, and
the lifetimes of $b$-flavoured hadrons. Determinations of $\alpha_s$
from $\Upsilon$ spectroscopy are briefly presented.
\end{abstract}

\vspace{1.5cm}

\centerline{\it To appear in the Proceedings of the}
\centerline{\it 17th International Conference on Lepton--Photon
Interactions}
\centerline{\it Beijing, China, August 1995}

\vspace{3.0cm}

\noindent
CERN-TH/95-307\\
November 1995

\end{titlepage}

\centerline{\normalsize\bf HEAVY FLAVOUR PHYSICS}
\vspace*{0.6cm}
\centerline{\footnotesize MATTHIAS NEUBERT}
\baselineskip=13pt
\centerline{\footnotesize\it Theory Division, CERN, CH-1211 Geneva
23, Switzerland}
\baselineskip=12pt
\centerline{\footnotesize E-mail: neubert@nxth04.cern.ch}
\vspace*{0.9cm}

\abstracts{
The current status of the theory and phenomenology of weak decays of
hadrons containing a heavy quark is reviewed. Exclusive semileptonic
and rare decays of $B$ mesons are discussed, as well as inclusive
decay rates, the semileptonic branching ratio of the $B$ meson, and
the lifetimes of $b$-flavoured hadrons. Determinations of $\alpha_s$
from $\Upsilon$ spectroscopy are briefly presented.
}

\normalsize\baselineskip=15pt
\setcounter{footnote}{0}
\renewcommand{\thefootnote}{\alph{footnote}}

\section{Introduction}

Studies of weak decays of heavy flavours play a key role in testing
the Standard Model and determining some of its parameters, which are
related to flavour physics. In these decays information about the
quark masses and the Cabibbo--Kobayashi--Maskawa (CKM) matrix can be
obtained. A precise measurement of these parameters is a prerequisite
for testing such intriguing phenomena as CP violation, and for
exploring new physics beyond the Standard Model. On the other hand,
weak decays of hadrons serve as a probe of that part of strong
interaction physics which is least understood: the confining forces
that bind quarks and gluons inside hadrons. In fact, the
phenomenology of hadronic weak decays is characterized by an
intricate interplay between the weak and strong interactions, which
has to be disentangled before any information about Standard Model
parameters can be extracted.

The recent experimental progress in heavy flavour physics has been
summarized in the talks by T.~Skwarnicki\cite{Tomasz},
J.~Kroll\cite{Joe} and S.L.~Wu\cite{Wu} at this Conference. Here I
will present the theoretical framework for a description and
interpretation of some of the data presented there. Since the
discovery of heavy-quark symmetry\cite{Shur}$^-$\cite{review} and the
establishment of the heavy-quark expansion\cite{EiFe}$^-$\cite{Fermi}
this field has flourished. It is, therefore, unavoidable that I have
to be selective and focus on few topics of particular interest. In
this selection I was guided mainly by the relevance of a subject to
current experiments. I apologize to all those authors whose work will
thus be omitted here. In particular, I will not be able to report on
theoretical progress in the areas of meson decay
constants\cite{dec1,dec2}, exclusive nonleptonic decays of $B$
mesons\cite{nonlep1,nonlep2}, and inclusive decay spectra in
semileptonic and rare $B$ decays\cite{spectra,moments}, although
there was a large activity devoted to these subjects. I will also
have to leave out some formal developments, such as the study of
renormalons in the heavy-quark effective
theory\cite{BBren}$^-$\cite{BaBB}.

This article is divided into two parts; the first covers exclusive
semileptonic and rare decays, the second is devoted to inclusive
decay rates and lifetimes. At the end, I will briefly discuss
extractions of the strong coupling constant $\alpha_s$ from
$\Upsilon$ spectroscopy.

\newpage
\vspace*{-0.5cm}
\section{Exclusive Semileptonic Decays}

\begin{figure}[htb]
   \epsfysize=3.5cm
   \centerline{\epsffile{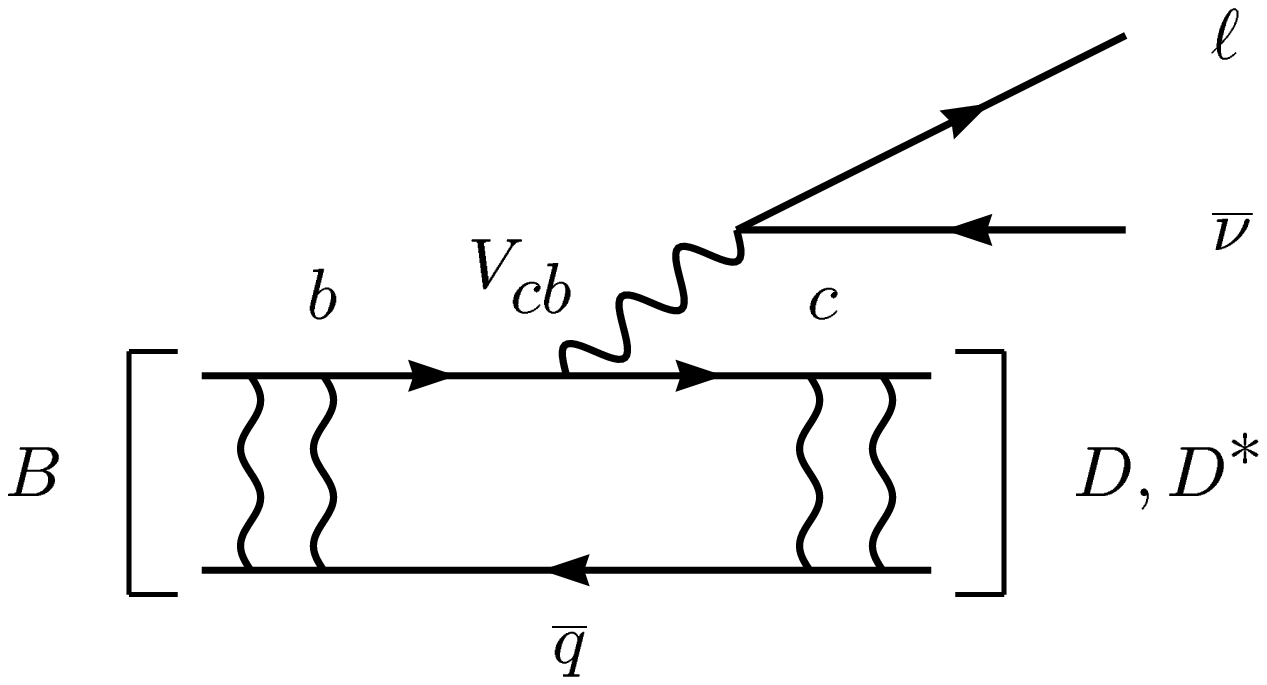}}
\fcaption{Semileptonic decay of a $B$ meson.}
\label{fig:1}
\end{figure}

Semileptonic decays of $B$ mesons have received a lot of attention in
recent years. The decay channel $B\to D^*\ell\,\bar\nu$ has the
largest branching fraction of all $B$-meson decay modes, and large
data samples have been collected by various experimental groups. From
the theoretical point of view, semileptonic decays are simple enough
to allow for a reliable, quantitative description. Yet, the analysis
of these decays provides much information about the strong forces
that bind the quarks and gluons into hadrons. Schematically, a
semileptonic decay process is shown in Fig.~\ref{fig:1}. The strength
of the $b\to c$ transition is governed by the element $V_{cb}$ of the
CKM matrix. The entries of this matrix are fundamental parameters of
the Standard Model. A primary goal of the study of semileptonic
decays of $B$ mesons is to extract with high precision the values of
$V_{cb}$ and $V_{ub}$. The problem is that the Standard Model
Lagrangian is formulated in terms of quark and gluon fields, whereas
the physical hadrons are bound states of these degrees of freedom.
Thus, an understanding of the transition from the quark to the hadron
world is necessary before the fundamental parameters can be extracted
from experimental data.

In the case of transitions between two heavy quarks, such as $b\to
c\,\ell\,\bar\nu$, heavy-quark symmetry helps to eliminate (or at
least reduce) the hadronic uncertainties in the theoretical
description. The physical picture underlying this symmetry is the
following\cite{Shur}$^-$\cite{review}: In a heavy-light bound state
such as a heavy meson or baryon, the typical momenta exchanged
between the heavy and light constituents are of order the confinement
scale $\Lambda$. The heavy quark is surrounded by a most complicated,
strongly interacting cloud of light quarks, antiquarks, and gluons.
However, the fact that $1/m_Q\ll 1/\Lambda$, i.e.\ the Compton
wavelength of the heavy quark is much smaller than the size of the
hadron, leads to simplifications. To resolve the quantum numbers of
the heavy quark would require a hard probe; soft gluons can only
resolve distances much larger than $1/m_Q$. Therefore, the light
degrees of freedom are blind to the flavour (mass) and spin of the
heavy quark; they only experience its colour field, which extends
over large distances because of confinement. It follows that, in the
$m_Q\to\infty$ limit, hadronic systems which differ only in the
flavour or spin quantum numbers of the heavy quark have the same
configuration of their light degrees of freedom. Although this
observation still does not allow us to calculate what this
configuration is, it provides relations between the properties of
such particles as the heavy mesons $B$, $D$, $B^*$ and $D^*$, or the
heavy baryons $\Lambda_b$ and $\Lambda_c$. These relations result
from new symmetries of the effective strong interactions of heavy
quarks at low energies\cite{Isgu}. The configuration of light degrees
of freedom in a hadron containing a single heavy quark with velocity
$v$ and spin $s$ does not change if this quark is replaced by another
heavy quark with different flavour or spin, but with the same
velocity. For $N_h$ heavy quark flavours, there is thus an SU$(2
N_h)$ spin-flavour symmetry. Most importantly, this symmetry relates
all hadronic form factors in semileptonic decays of the type $B\to
D\,\ell\,\bar\nu$ and $B\to D^*\ell\,\bar\nu$ to a single universal
form factor, the Isgur--Wise function, and fixes the normalization of
this function at maximum $q^2$ (corresponding to zero recoil or equal
meson velocities). Heavy-quark symmetry is an approximate symmetry,
and corrections of order $\alpha_s(m_Q)$ or $\Lambda/m_Q$ arise since
the quark masses are not infinite. A systematic framework for
analyzing them is provided by the heavy-quark effective theory
(HQET)\cite{Geor}$^-$\cite{Luke}.

\subsection{Determination of\/ $|\,V_{cb}|$}

A model-independent determination of $|\,V_{cb}|$ based on
heavy-quark symmetry can be obtained by measuring the recoil spectrum
of $D^*$ mesons produced in $B\to D^*\ell\,\bar\nu$ decays\cite{Vcb}.
In terms of the variable
\begin{equation}\label{wq2rel}
   w = v_B\cdot v_{D^*} = {E_{D^*}\over m_{D^*}}
   = {m_B^2 + m_{D^*}^2 - q^2\over 2 m_B m_{D^*}} \,,
\end{equation}
the differential decay rate reads
\begin{eqnarray}
   {{\rm d}\Gamma\over{\rm d}w}
   &=& {G_F^2\over 48\pi^3}\,(m_B-m_{D^*})^2\,m_{D^*}^3
    \sqrt{w^2-1}\,(w+1)^2 \nonumber\\
   &&\mbox{}\times \bigg[ 1 + {4w\over w+1}\,
    {m_B^2-2w\,m_B m_{D^*} + m_{D^*}^2\over(m_B - m_{D^*})^2}
    \bigg]\,|\,V_{cb}|^2\,{\cal{F}}^2(w) \,.
\end{eqnarray}
The hadronic form factor ${\cal F}(w)$ coincides with the Isgur--Wise
function up to symme\-try-breaking corrections of order
$\alpha_s(m_Q)$ and $\Lambda/m_Q$. The idea is to measure the product
$|\,V_{cb}|\,{\cal F}(w)$ as a function of $w$, and to extract
$|\,V_{cb}|$ from an extrapolation of the data to the zero-recoil
point $w=1$, where the $B$ and the $D^*$ mesons have a common rest
frame. At this kinematic point, heavy-quark symmetry helps to
calculate the normalization ${\cal F}(1)$ with small and controlled
theoretical errors, so that the determination of $|\,V_{cb}|$ becomes
model independent. Since the range of $w$ values accessible in this
decay is rather small ($1<w<1.5$), the extrapolation can be done
using an expansion around $w=1$,
\begin{equation}\label{Fexp}
   {\cal F}(w) = {\cal F}(1)\,\Big\{ 1 - \widehat\varrho^2\,(w-1)
   + \widehat c\,(w-1)^2 + \dots \Big\} \,.
\end{equation}
Usually a linear form of the form factor is assumed, and the slope
$\widehat\varrho^2$ is treated as a parameter.

Measurements of the recoil spectrum have been performed first by the
ARGUS\cite{ARGVcb} and CLEO\cite{CLEOVcb} Collaborations in
experiments operating at the $\Upsilon(4s)$ resonance, and more
recently by the ALEPH\cite{ALEVcb} and DELPHI\cite{DELVcb}
Collaborations at LEP. These measurements have been discussed in
detail by T.~Skwarnicki\cite{Tomasz} at this Conference. The weighted
average of the results is
\begin{eqnarray}\label{VcbF}
   |\,V_{cb}|\,{\cal F}(1) &=& (35.1\pm 1.7_{-0.0}^{+1.4})
   \times 10^{-3} \,, \nonumber\\
   \widehat\varrho^2 &=& 0.87\pm 0.16 \,.
\end{eqnarray}
The effect of a positive curvature of the form factor has been
investigated by Stone\cite{Stone}, who finds that the value of
$|\,V_{cb}|\,{\cal F}(1)$ may change by up to $+4\%$. This
uncertainty is included by the second error quoted above.

\subsubsection{Calculations of ${\cal F}(1)$}

Heavy-quark symmetry implies that the general structure of the
symmetry-break\-ing corrections to the form factor at zero recoil
must be of the form\cite{Vcb}
\begin{equation}
   {\cal F}(1) = \eta_A\,\bigg( 1 + 0\cdot {\Lambda\over m_Q}
   + c_2\,{\Lambda^2\over m_Q^2} + \dots \bigg)
   = \eta_A\,(1+\delta_{1/m^2}) \,,
\end{equation}
where $\eta_A$ is a short-distance correction arising from a finite
renormalization of the flavour-changing axial current at zero recoil,
and $\delta_{1/m^2}$ parametrizes second-order (and higher) power
corrections. The absence of first-order power corrections at zero
recoil is a consequence of the Luke theorem\cite{Luke}, which is the
analogue of the Ademollo--Gatto theorem\cite{AGTh} for heavy-quark
symmetry.

The one-loop expression for $\eta_A$ is known since a long
time\cite{Pasc,Volo,QCD1}:
\begin{equation}
   \eta_A = 1 + {\alpha_s(M)\over\pi}\,\bigg(
   {m_b+m_c\over m_b-m_c}\,\ln{m_b\over m_c} - {8\over 3} \bigg)
   \simeq 0.96 \,.
\end{equation}
The scale $M$ in the running coupling constant can be fixed by
adopting the prescription of Brodsky, Lepage and Mackenzie\cite{BLM},
where it is identified with the average virtuality of the gluon in
the one-loop diagrams that contribute to $\eta_A$. If $\alpha_s(M)$
is defined in the $\overline{\mbox{\sc ms}}$ scheme, the result
is\cite{etaVA} $M\simeq 0.51\sqrt{m_c m_b}$. Several estimates of
higher-order corrections to $\eta_A$ have been discussed. A
renormalization-group resummation of mass logarithms of the type
$(\alpha_s\ln m_b/m_c)^n$, $\alpha_s(\alpha_s\ln m_b/m_c)^n$ and
$m_c/m_b(\alpha_s\ln m_b/m_c)^n$ leads to\cite{PoWi}$^-$\cite{QCD2}
$\eta_A\simeq 0.985$. On the other hand, a resummation of
renormalon-chain contributions of the form $\beta_0^{n-1}\alpha_s^n$,
where $\beta_0=11-\frac{2}{3}n_f$ is the first coefficient of the
$\beta$-function, gives\cite{flow} $\eta_A\simeq 0.945$. Using these
partial resummations to estimate the uncertainty, I quote
\begin{equation}
   \eta_A = 0.965\pm 0.020 \,.
\end{equation}
The accuracy of this result could be improved with an exact two-loop
calculation.

An analysis of the power corrections $\delta_{1/m^2}$ is more
difficult, since it cannot rely on perturbation theory. Three
approaches have been discussed: in the ``exclusive approach'', all
$1/m_Q^2$ operators in the HQET are classified and their matrix
elements estimated\cite{FaNe,Mann}, leading to $\delta_{1/m^2}=-(3\pm
2)\%$; the ``inclusive approach'' has been used to derive the
bound\cite{Shif} $\delta_{1/m^2}<-3\%$, and to estimate that
$\delta_{1/m^2}=-(7\pm 3)\%$; the ``hybrid approach'' combines the
virtues of the former two to obtain a more restrictive lower bound on
$\delta_{1/m^2}$. The result is\cite{Vcbnew}
\begin{equation}
   \delta_{1/m^2} = - 0.055\pm 0.025 \,,
\end{equation}
which is consistent with previous estimates. To obtain a more precise
prediction, one should attempt to calculate this quantity using
lattice simulations of QCD.

Combining the above results, adding the theoretical errors linearly
to be conservative, gives
\begin{equation}\label{F1}
   {\cal F}(1) = 0.91\pm 0.04
\end{equation}
for the normalization of the hadronic form factor at zero recoil.
This can be used to extract from the experimental result
(\ref{VcbF}) the model-independent value
\begin{equation}\label{Vcbexc}
   |\,V_{cb}| = (38.6_{-1.9}^{+2.4}\phantom{\Big|}_{\rm exp}
   \pm 1.7_{\rm th})\times 10^{-3} \,.
\end{equation}
After $|\,V_{ud}|$ and $|\,V_{us}|$, this is now the third-best known
entry in the CKM matrix.

\subsubsection{Bounds and predictions for $\widehat\varrho^2$}

\begin{figure}[htb]
   \epsfysize=6cm
   \centerline{\epsffile{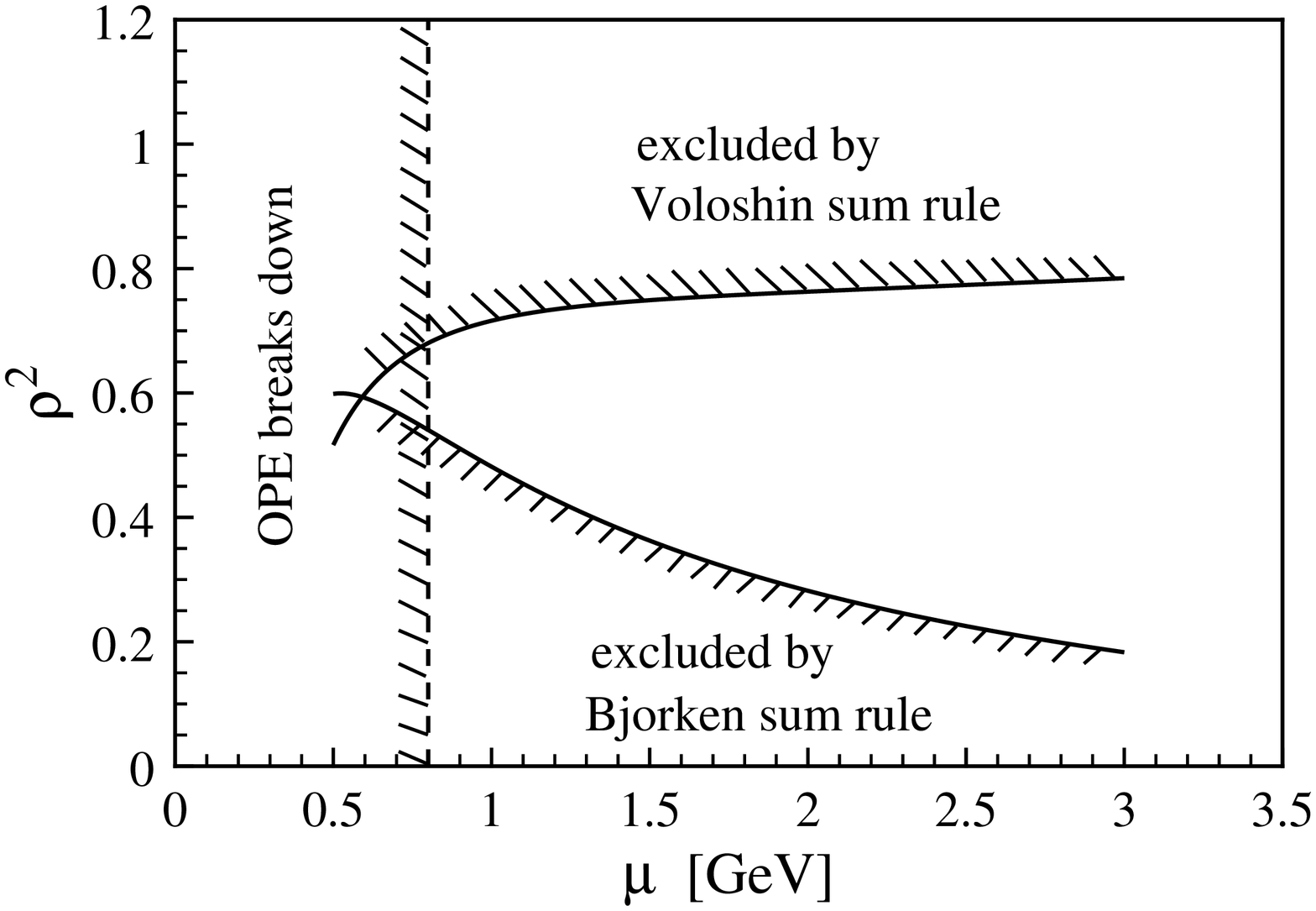}}
\fcaption{Bounds for the slope parameter $\varrho^2$ following from
the Bjorken and Voloshin sum rules.}
\label{fig:2}
\end{figure}

The slope parameter $\widehat\varrho^2$ in the expansion of the
physical form factor in (\ref{Fexp}) differs from the slope parameter
$\varrho^2$ of the universal Isgur--Wise function by corrections that
violate the heavy-quark symmetry. The short-distance corrections have
been calculated, with the result\cite{Vcbnew}
\begin{equation}\label{rhorel}
   \widehat\varrho^2 = \varrho^2 + (0.16\pm 0.02) + O(1/m_Q) \,.
\end{equation}
The slope of the Isgur--Wise function is constrained by sum rules,
which relate the inclusive decay rate of the $B$ meson to a sum over
exclusive channels. At lowest order, Bjorken and Voloshin have
derived two such sum rules, which imply the
bounds\cite{Bjsum}$^-$\cite{Volsum}
\begin{equation}
   {1\over 4} < \varrho^2 < {1\over 4} + {\bar\Lambda\over 2 E_0}
   \simeq 0.8 \,,
\end{equation}
where $\bar\Lambda=m_B-m_b$, and $E_0=m_{B^{**}}-m_B$. Corrections to
this result can be calculated in a systematic way using the Operator
Product Expansion (OPE), where one introduces a momentum scale
$\mu\sim\mbox{few}\times\Lambda$ chosen large enough so that
$\alpha_s(\mu)$ and power corrections of order $(\Lambda/\mu)^n$ are
small, but otherwise as small as possible to suppress contributions
from excited states\cite{GrKo}. The result is\cite{KoNe}
$\varrho_{\rm min}^2(\mu) < \varrho^2 < \varrho_{\rm max}^2(\mu)$,
where the boundary values are shown in Fig.~\ref{fig:2} as a function
of the scale $\mu$. Assuming that the OPE works down to values
$\mu\simeq 0.8$~GeV, one obtains rather tight bounds for the slope
parameters:
\begin{eqnarray}\label{rhobounds}
   0.5 &< \varrho^2 &< 0.8 \,, \nonumber\\
   0.5 &< \widehat\varrho^2 &< 1.1 \,.
\end{eqnarray}
The allowed region for $\widehat\varrho^2$ has been increased in
order to account for the unknown $1/m_Q$ corrections in the relation
(\ref{rhorel}). The experimental result given in (\ref{VcbF}) falls
inside this region.

These bounds compare well with theoretical calculations of the slope
parameters. QCD sum rules have been used to calculate the slope of
the Isgur--Wise function; the results obtained by various authors are
$\varrho^2 = 0.84\pm 0.02$ (Bagan {\it et al}.\cite{Baga}), $0.7\pm
0.1$ (Neubert\cite{twoloop}), $0.70\pm 0.25$ (Blok and
Shifman\cite{BlSh}), and $1.00\pm 0.02$ (Narison\cite{Nari}). The
UKQCD Collaboration has presented a lattice calculation of the slope
of the form factor ${\cal F}(w)$, yielding\cite{Lattrho}
$\widehat\varrho^2 = 0.9_{-0.3-0.2}^{+0.2+0.4}$. I stress that the
sum rule bounds in (\ref{rhobounds}) are largely model independent;
model calculations in strong disagreement with these bounds should be
discarded.

\subsubsection{Analyticity bounds and correlations between
$\widehat\varrho^2$ and $\widehat c$}

A model-independent method of constraining the $q^2$ dependence of
form factors using analyticity properties of QCD spectral functions
and unitarity was proposed some time ago\cite{Bour}. This method has
been applied to the elastic form factor of the $B$
meson\cite{Taro}$^-$\cite{Boyd1}, which is related by heavy-quark
symmetry to the Isgur--Wise function. It has also been applied
directly to the form factors of interest for $B\to
D^{(*)}\ell\,\bar\nu$ decays\cite{Boyd2}. Thereby, bounds have been
derived for the slope and curvature of the function ${\cal F}(w)$ in
(\ref{Fexp}). These bounds are rather weak, however, due to the
presence of $B_c$ poles below threshold. The lack of information
about the residues of these poles reduced considerably the
constraining power of the method.

The problem of sub-threshold poles, which weaken the analyticity
bounds, can be avoided by using heavy-quark symmetry\cite{CaNe}.
Instead of relying on model-dependent predictions about the
properties of $B_c$ mesons, one can identify a specific $B\to D$ form
factor which does not receive contributions from the ground-state
$B_c$ poles. Strong model-independent constraints on the slope and
curvature of this form factor can be derived, and heavy-quark
symmetry can be used to relate this form factor to the function
${\cal F}(w)$ describing $B\to D^*\ell\,\bar\nu$ decays. These
relations receive symmetry-breaking corrections, which can however be
estimated and turn out to weaken the bounds only softly.

For a more detailed discussion of the results of these analyses, I
refer to the original literature\cite{Boyd2,CaNe}.

\subsection{Measurement of $B\to D^*\ell\,\bar\nu$ form factors}

If the lepton mass is neglected, the differential decay distributions
in $B\to D^*\ell\,\bar\nu$ decays can be parametrized by three
helicity amplitudes, or equivalently by three independent
combinations of form factors. It has been suggested that a good
choice for such three quantities should be inspired by the
heavy-quark expansion\cite{review,subl}. One thus defines a form
factor $h_{A1}(w)$, which up to symmetry-breaking corrections
coincides with the Isgur--Wise function, and two form factor ratios
\begin{eqnarray}
   R_1(w) &=& \bigg[ 1 - {q^2\over(m_B+m_{D^*})^2} \bigg]\,
    {V(q^2)\over A_1(q^2)} \,, \nonumber\\
   R_2(w) &=& \bigg[ 1 - {q^2\over(m_B+m_{D^*})^2} \bigg]\,
    {A_2(q^2)\over A_1(q^2)} \,.
\end{eqnarray}
The relation between $w$ and $q^2$ has been given in (\ref{wq2rel}).
This definition is such that in the heavy-quark limit
$R_1(w)=R_2(w)=1$ independently of $w$.

To extract the functions $h_{A1}(w)$, $R_1(w)$ and $R_2(w)$ from
experimental data is a complicated task. However, HQET-based
calculations suggest that the $w$-dependence of the form factor
ratios, which is induced by symmetry-breaking effects, is rather
mild\cite{subl}. Moreover, the form factor $h_{A1}(w)$ is expected to
have a nearly linear shape over the accessible $w$ range. This
motivates to introduce three parameters $\varrho_{A1}^2$, $R_1$ and
$R_2$ by
\begin{eqnarray}
   h_{A1}(w) &=& {\cal F}(1)\,\Big\{ 1 - \varrho_{A1}^2 (w-1)
    + O[(w-1)^2] \Big\} \,, \nonumber\\
   R_1(w) &=& R_1\,\Big\{ 1 + O(w-1) \Big\} \,, \\
   R_2(w) &=& R_2\,\Big\{ 1 + O(w-1) \Big\} \,, \nonumber
\end{eqnarray}
where ${\cal F}(1)=0.91\pm 0.04$ from (\ref{F1}). The CLEO
Collaboration has extracted these three parameters from a joint
analysis of the angular distributions in $B\to D^*\ell\,\bar\nu$
decays\cite{CLEff}. The result is:
\begin{eqnarray}
   \varrho_{A1}^2 &=& 0.91\pm 0.15\pm 0.06 \,, \nonumber\\
   R_1 &=& 1.18\pm 0.30\pm 0.12 \,, \\
   R_2 &=& 0.71\pm 0.22\pm 0.07 \,. \nonumber
\end{eqnarray}
Using the HQET, one obtains an essentially
model-independent prediction for the symmetry-breaking corrections to
$R_1$, whereas the corrections to $R_2$ are more model dependent. To
good approximation\cite{review}
\begin{eqnarray}
   R_1 &\simeq& 1 + {4\alpha_s(m_c)\over 3\pi}
    + {\bar\Lambda\over 2 m_c}\simeq 1.3\pm 0.1 \,, \nonumber\\
   R_2 &\simeq& 1 - \kappa\,{\bar\Lambda\over 2 m_c}
    \simeq 0.8\pm 0.2 \,,
\end{eqnarray}
with $\kappa\simeq 1$ from QCD sum rules\cite{subl}. A quark-model
calculation of $R_1$ and $R_2$ gives similar results\cite{ClWa}:
$R_1\simeq 1.15$ and $R_2\simeq 0.91$. The experimental data confirm
the theoretical prediction that $R_1>1$ and $R_2<1$, although the
errors are still large.

There is a model-independent relation between the three parameters
determined from the joined angular analysis and the slope parameter
$\widehat\varrho^2$ extracted from the semileptonic spectrum. It
reads\cite{Vcbnew}
\begin{equation}
   \varrho_{A1}^2 - \widehat\varrho^2 = {1\over 6}\,(R_1^2-1)
   + {m_B\over 3(m_B-m_{D^*})}\,(1-R_2) \,.
\end{equation}
The CLEO data give $0.07\pm 0.20$ for the difference of the slope
parameters on the left-hand side, and $0.22\pm 0.18$ for the
right-hand side. Both values are compatible within errors.

In my opinion the results of this analysis are very encouraging.
Within errors, the experiment confirms the HQET predictions, starting
to test them at the level of symmetry-breaking corrections.

\subsection{Decays to charmless final states}

Very recently, the CLEO Collaboration has reported a first signal for
exclusive semileptonic decays of $B$ mesons into charmless final
states in the decay modes $B\to\pi\,\ell\,\bar\nu$ and
$B\to\rho\,\ell\,\bar\nu$. The underlying quark process for these
transitions is $b\to u\,\ell\,\bar\nu$. Thus, these decays provide
information on the strength of the CKM matrix element $V_{ub}$. The
observed branching fractions are\cite{CLEVub}:
\begin{eqnarray}\label{CLEOVub}
   \mbox{B}(B\to\pi\,\ell\,\bar\nu) &=& \cases{
    (1.34\pm 0.45)\times 10^{-4} ;&ISGW, \cr
    (1.63\pm 0.57)\times 10^{-4} ;&BSW, \cr} \nonumber\\
   && \\
   \mbox{B}(B\to\rho\,\ell\,\bar\nu) &=& \cases{
    (2.28_{-0.83}^{+0.69})\times 10^{-4} ;&ISGW, \cr
    (3.88_{-1.39}^{+1.15})\times 10^{-4} ;&BSW. \cr} \nonumber
\end{eqnarray}
There is a significant model dependence in the simulation of the
reconstruction efficiencies, for which the models of Isgur, Scora,
Grinstein and Wise\cite{ISGW} (ISGW) and Bauer, Stech and
Wirbel\cite{BSW} (BSW) have been used.

The theoretical description of these heavy-to-light ($b\to u$) decays
is more model dependent than that for heavy-to-heavy ($b\to c$)
transitions, because heavy-quark symmetry does not help to determine
the relevant hadronic form factors. A variety of calculations for
such form factors exists, based on QCD sum rules, lattice gauge
theory, perturbative QCD, or quark models. In Table~\ref{tab:Vub}, I
give a summary of values extracted for the ratio $|\,V_{ub}/V_{cb}|$
from a selection of such calculations. Clearly, some approaches are
more consistent than others in that the extracted values are
compatible for the two decay modes. With few exceptions, the results
lie in the range
\begin{equation}
   \left| {V_{ub}\over V_{cb}} \right|_{\rm excl}
   = 0.06\mbox{--}0.11 \,,
\end{equation}
which is in good agreement with the measurement of $|\,V_{ub}|$
obtained from endpoint region of the lepton spectrum in inclusive
semileptonic decays\cite{Vubin1,Vubin2}:
\begin{equation}
   \left| {V_{ub}\over V_{cb}} \right|_{\rm incl}
   = 0.08\pm 0.01_{\rm exp}\pm 0.02_{\rm th} \,.
\end{equation}

\begin{table}[htb]
\tcaption{Values for $|\,V_{ub}/V_{cb}|$ extracted from the CLEO
measurement of exclusive semileptonic $B$ decays into charmless final
states, taking $|\,V_{cb}|=0.040$. An average over the experimental
results in (\ref{CLEOVub}) is used for all except the ISGW and BSW
models, where the numbers corresponding to these models are used. The
first error quoted is experimental, the second (when available) is
theoretical. }
\label{tab:Vub}
\vspace{0.2cm}
\small
\centerline{\begin{tabular}{|c|c|c|c|}\hline\hline
\rule[-0.2cm]{0cm}{0.65cm} Method & Reference
 & $B\to\pi\,\ell\,\bar\nu$ & $B\to\rho\,\ell\,\bar\nu$ \\
\hline
\rule[-0.2cm]{0cm}{0.65cm} QCD sum rules & Narison\protect\cite{SNar}
 & $0.159\pm 0.019\pm 0.001$ & $0.066_{-0.009}^{+0.007}\pm 0.003$ \\
\rule[-0.2cm]{0cm}{0.65cm} & Ball\protect\cite{PBal}
 & $0.105\pm 0.013\pm 0.011$ & $0.094_{-0.012}^{+0.010}\pm 0.016$ \\
\rule[-0.2cm]{0cm}{0.65cm} & Yang, Hwang\cite{YaHw}
 & $0.102\pm 0.012_{-0.013}^{+0.015}$ &
 $0.184_{-0.024-0.015}^{+0.020+0.027}$ \\
\hline
\rule[-0.2cm]{0cm}{0.65cm} lattice QCD & UKQCD\cite{UKQCDVub}
 & $0.103\pm 0.012_{-0.010}^{+0.012}$ & --- \\
\rule[-0.2cm]{0cm}{0.65cm} & APE\cite{APEVub}
 & $0.084\pm 0.010\pm 0.021$ & --- \\
\hline
\rule[-0.2cm]{0cm}{0.65cm} pQCD & Li, Yu\cite{LiYu}
 & $0.054\pm 0.006$ & --- \\
\hline
\rule[-0.2cm]{0cm}{0.65cm} quark models & BSW\cite{BSW}
 & $0.093\pm 0.016$ & $0.076_{-0.014}^{+0.011}$ \\
\rule[-0.2cm]{0cm}{0.65cm} & KS\cite{KS} & $0.088\pm 0.011$
 & $0.056_{-0.007}^{+0.006}$ \\
\rule[-0.2cm]{0cm}{0.65cm} & ISGW2\cite{ISGW2} & $0.074\pm 0.012$
 & $0.079_{-0.014}^{+0.012}$ \\
\hline\hline
\end{tabular}}
\end{table}

Clearly, this is only the first step towards a more reliable
determination of $|\,V_{ub}|$; yet, with the discovery of exclusive
$b\to u$ transitions an important milestone has been met. Efforts
must now concentrate on more reliable methods to determine the form
factors for heavy-to-light transitions. Some new ideas in this
direction have been discussed recently. They are based on lattice
calculations\cite{Flynn}, analyticity constraints\cite{Boyd1,Lell},
or variants of the form-factor relations for heavy-to-heavy
transitions\cite{Stech}.

\section{Exclusive rare radiative decays}

\begin{figure}[htb]
   \epsfysize=3.0cm
   \centerline{\epsffile{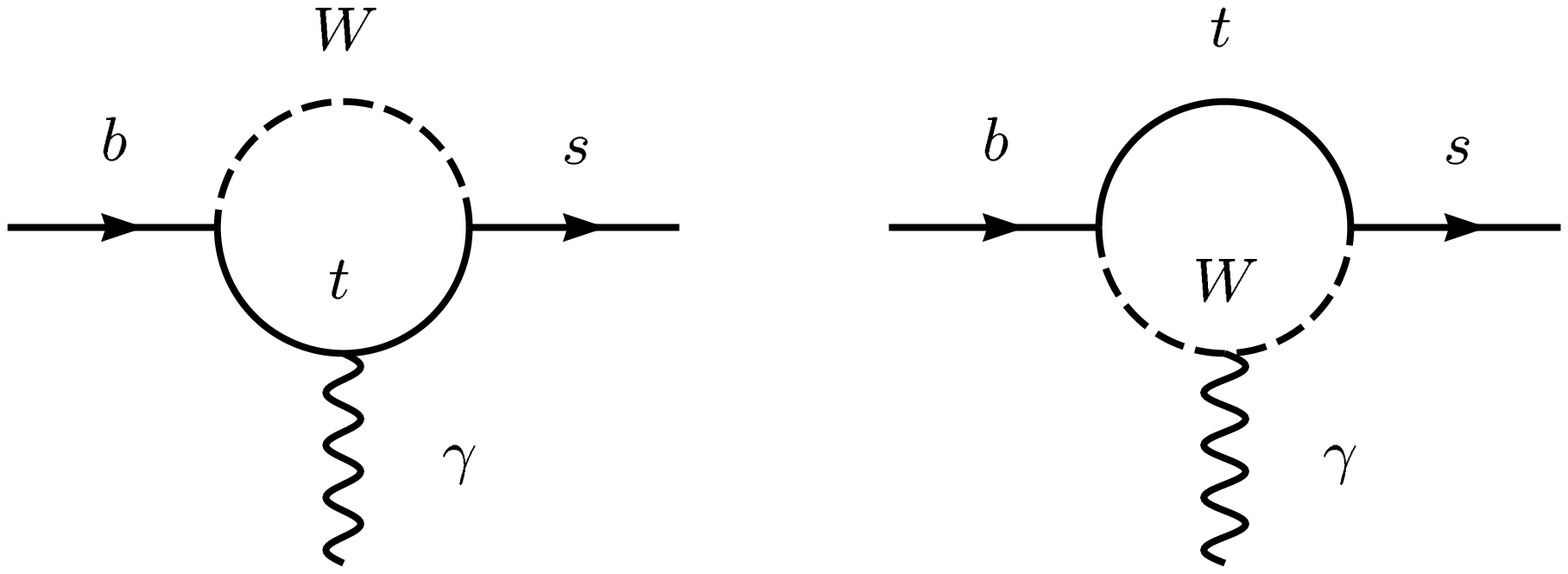}}
\fcaption{Penguin diagrams for the quark transition $b\to s\gamma$.}
\label{fig:penguin}
\end{figure}

Rare decays of $B$ mesons play an important role in testing the
Standard Model, as they are sensitive probes of new physics at high
energy scales. On the quark level, rare decays involve
flavour-changing neutral currents such as $b\to s\gamma$ or $b\to
d\,\ell^+\ell^-$. In the Standard Model they are forbidden at the
tree level, but can proceed at the one-loop level through penguin or
box diagrams, see Fig.~\ref{fig:penguin}.

The effective Hamiltonian describing the rare radiative decay $b\to
s\gamma$ is
\begin{equation}
   {\cal H}_{\rm eff} = - {G_F\over\sqrt{2}}\,V_{ts}^*\,V_{tb}\,
   c_7(\mu)\,{e\,m_b\over 8\pi^2}\,\bar s\sigma_{\mu\nu}
   (1+\gamma_5)b\,F^{\mu\nu} + \dots \,.
\end{equation}
The Wilson coefficient $c_7(\mu)$ contains the short-distance physics
of the heavy particles in the loop ($t$ and $W$ in the Standard
Model). Its value is sensitive to new physics, such as the existence
of charged Higgs bosons, which can in principle be probed by
measuring the inclusive decay rates for $B\to X_{s,d}\gamma$. The
uncertainty in the calculation of $c_7(\mu)$ in the Standard Model
is still of order\cite{c7,Bura} $\pm 15\%$, however, reducing
significantly the constraining power of such measurements.

The study of exclusive rare decays focuses on ratios such as
\begin{equation}
   R_{K^*} = {\Gamma(B\to K^*\gamma)\over\Gamma(B\to X_s\gamma)}
   \,,\qquad
   R_\rho = {\Gamma(B\to\rho\,\gamma)\over\Gamma(B\to X_d\gamma)} \,,
\end{equation}
which are no longer sensitive to new physics (since the coefficient
$c_7(\mu)$ cancels out), but test some strong interaction matrix
elements. The measurement reported by the CLEO
Collaboration\cite{CLRK},
\begin{equation}
   R_{K^*} = (19\pm 7\pm 4)\% \,,
\end{equation}
can be confronted with theoretical predictions, which have however a
wide spread. QCD sum-rule calculations lead to results in the
range\cite{Cola}$^-$\cite{SNar2} $R_{K^*}=(17\pm 5)\%$, while quark
model predictions range between\cite{RKmodels} 4\% and 30\%. Lattice
simulations of the relevant form factors have been performed over a
limited range in $q^2$ only, and the results for $R_{K^*}$ depend
rather strongly on the assumption about the $q^2$ dependence outside
this region. Studies of the various
groups\cite{UKQCDVub,BHS}$^-$\cite{LANL} have led to values between
5\% and 35\%. More work is needed to reduce the systematic
uncertainties in these calculations.

\begin{figure}[htb]
   \epsfysize=3.5cm
   \centerline{\epsffile{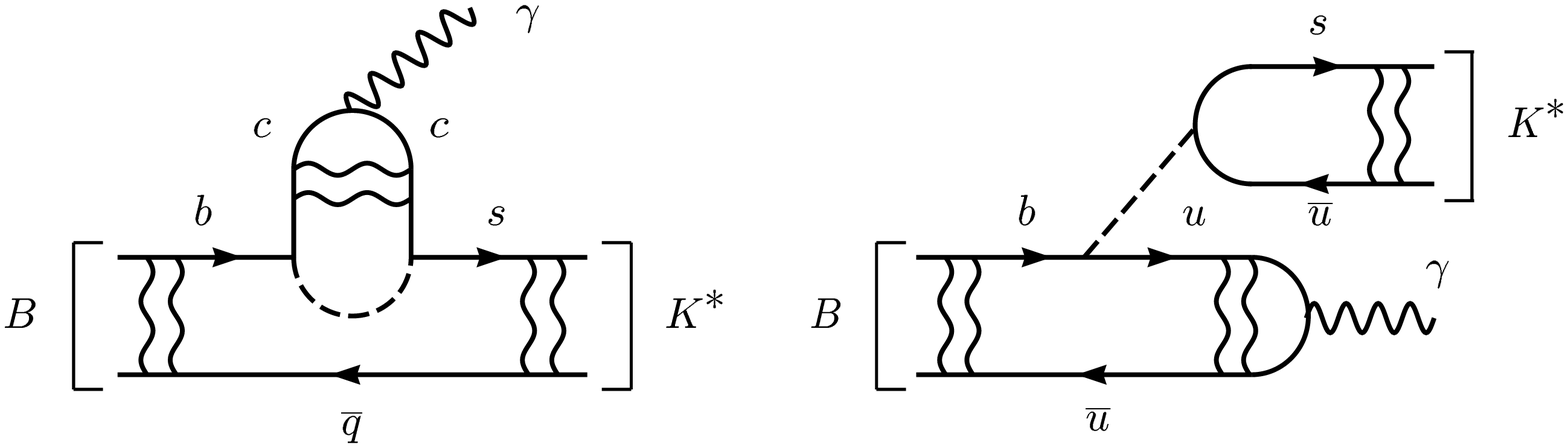}}
\fcaption{Long-distance contributions to rare radiative decays.}
\label{fig:longdist}
\end{figure}

In the theoretical analyses described above, it is assumed that $B\to
K^*\gamma$ decays are short-distance dominated. This assumption has
been questioned recently by Atwood, Blok and Soni\cite{ABS}, who
pointed out the possibility of large long-distance contributions in
the decay $B\to K^*\gamma$, and even more so in the decay
$B\to\rho\,\gamma$. Examples of such long-distance contributions are
shown in Fig.~\ref{fig:longdist}. The first graph shows a
``long-distance penguin'' diagram, in which the $c$-quark in the loop
is close to its mass shell. The $c\bar c$ pair forms a virtual vector
meson state $\psi^*$, which then decays into a photon. The second
graph shows the ``weak annihilation'' of the quark and antiquark in
the $B$ meson. For $B\to K^*\gamma$, this process is CKM suppressed
with respect to the penguin diagram, but this suppression is not
operative for $B\to\rho\,\gamma$.

Estimates of these long-distance contributions are difficult and
currently contro\-ver\-sial\cite{Golo}$^-$\cite{Khod}. For $B\to
K^*\gamma$, predictions for the ratio of the long- and short-distance
amplitudes, $|A_{\rm ld}/A_{\rm sd}|$, range from 15--50\% to $<
10\%$. In $B^-\to\rho^-\gamma$ decays, most authors expect
long-distance effects at a level of 10--30\%, whereas the effects are
much smaller, $\sim 1$--10\%, in the neutral channel
$B^0\to\rho^0\gamma$. Further investigation of this important subject
is necessary before a conclusion can be drawn. A clarification of
this issue is also important with regard to a determination of the
ratio of CKM elements $|\,V_{td}/V_{ts}|$ from the comparison of the
decay rates for $B\to\rho\,\gamma$ and $B\to K^*\gamma$.

\section{Inclusive decay rates}

Inclusive decay rates determine the probability of the decay of a
particle into the sum of all possible final states with given quantum
numbers. The theoretical framework to describe inclusive decays of
heavy flavours is provided by the $1/m_Q$
expansion\cite{Chay}$^-$\cite{Fermi}, which is a ``Minkowskian
version'' of the OPE. This means that the theoretical treatment of
inclusive rates has a solid foundation in QCD, however with one
assumption: that of quark--hadron duality. In the description of
semileptonic decays (e.g.\ $B\to\ell\,\bar\nu+\mbox{hadrons}$), where
the integration over the lepton and neutrino phase space provides a
``smearing'' over the invariant hadronic mass, so-called ``global''
duality is needed\cite{PQW}, whereas the treatment of nonleptonic
decays (e.g.\ $B\to\mbox{hadrons}$), for which the total hadronic
mass is fixed, requires the stronger assumption of local duality. It
is important to stress that quark--hadron duality cannot be derived
from QCD, although it is a common assumption in QCD phenomenology.

The main results of the $1/m_Q$ expansion for inclusive decays are
that the free quark decay (i.e.\ the parton model) provides the first
term in a systematic $1/m_Q$ expansion, and the nonperturbative
corrections to it are suppressed by (at least) two powers of the
heavy quark mass, i.e.\ they are of relative order $(\Lambda/m_Q)^2$.
The generic expression of any inclusive decay rate of a hadron $H_Q$
containing a heavy quark into some final state with quantum numbers
$f$ is of the form\cite{Bigi}$^-$\cite{MaWe,BloS,liferef}
\begin{eqnarray}\label{generic}
   \Gamma(H_Q\to X_f) &=& {G_F^2 m_Q^5\over 192\pi^3}\,
    |\mbox{KM}|^2\,\bigg\{ c_3^f\,\bigg( 1
    - {\langle\bar Q(i\vec D)^2 Q\rangle_H\over 2 m_Q^2} \bigg)
    \nonumber\\
   &&\mbox{}+ c_5^f\,{\langle\bar Q g_s\sigma_{\mu\nu} G^{\mu\nu}
    Q\rangle_H\over m_Q^2} + \sum_i c_{6,i}^f\,
    {\langle\bar Q\Gamma_i q\bar q\Gamma_i Q\rangle_H\over m_Q^3}
    + \dots \bigg\} \,,
\end{eqnarray}
where $|\mbox{KM}|$ is a combination of CKM matrix elements, $c_n^f$
are calculable coefficient functions, and $\langle O_n\rangle_H$ are
the (normalized) forward matrix elements of local operators between
$H_Q$ states. The matrix elements of the dimension-five operators
are\cite{FaNe}
\begin{eqnarray}
   \langle\bar Q(i\vec D)^2 Q\rangle_H &=& - \lambda_1
    = \mu_\pi^2 \,, \nonumber\\
   \langle\bar Q g_s\sigma_{\mu\nu} G^{\mu\nu} Q\rangle_H
   &=& 2 d_H\lambda_2 \,,
\end{eqnarray}
where $d_P=3$, $d_V=-1$ and $\lambda_2=\frac{1}{4}(m_{B^*}^2 -
m_B^2)\simeq 0.12$~GeV$^2$ for the ground state pseudoscalar ($P_Q$)
and vector ($V_Q$) mesons, and $d_\Lambda=0$ for the $\Lambda_Q$
baryon. The matrix element of the ``kinetic energy operator'',
$\mu_\pi^2=-\lambda_1$, has been estimated by several
authors\cite{lam1}$^-$\cite{virial}; below I shall use the value
$-\lambda_1=(0.4\pm 0.2)$~GeV$^2$ with a conservative error. Meson
matrix elements of the dimension-six operator in (\ref{generic}) can
be related, in the vacuum saturation approximation\cite{SVZ}, to the
decay constant $f_B$ of the $B$ meson. I shall now discuss the most
important applications of this general formalism to inclusive decays
of $b$-flavoured mesons and baryons.

\subsection{Determination of\/ $|\,V_{cb}|$ from inclusive
semileptonic decays}

The extraction of $|\,V_{cb}|$ from the inclusive semileptonic decay
rate of the $B$ meson is based on the
expression\cite{Bigi}$^-$\cite{MaWe}
\begin{eqnarray}\label{Gamsl}
   \Gamma(B\to X_c\ell\,\bar\nu) &=& {G_F^2 m_b^5\over 192\pi^3}\,
    |\,V_{cb}|^2\,\bigg\{ \bigg( 1
    + {\lambda_1+3\lambda_2\over 2 m_b^2} \bigg)\,f(m_c/m_b)
    \nonumber\\
   &&\mbox{}- {6\lambda_2\over m_b^2}\,\bigg( 1 - {m_c^2\over m_b^2}
    \bigg)^4 + {\alpha_s(M)\over\pi}\,g(m_c/m_b) + \dots \bigg\} \,,
\end{eqnarray}
where $m_b$ is the pole mass of the $b$ quark (defined to a given
order in perturbation theory), and $f(x)$ and $g(x)$ are phase space
functions given elsewhere\cite{fgrefs}. The theoretical uncertainties
in this determination of $|\,V_{cb}|$ are quite different from the
ones entering the analysis of exclusive decays. In inclusive decays
there appear the quark masses rather than the meson masses. Moreover,
the theoretical description relies on the assumption of global
quark--hadron duality, which is not necessary for exclusive decays. I
will now discuss the theoretical uncertainties in detail.

\subsubsection{Nonperturbative corrections}

The nonperturbative corrections are very small; with
$-\lambda_1=(0.4\pm 0.2)$~GeV$^2$ and $\lambda_2=0.12$~GeV$^2$, one
finds a reduction of the parton model rate by $-(4.2\pm 0.5)\%$. The
uncertainty in this number is below 1\% and thus completely
negligible.

\subsubsection{Dependence on quark masses}

\begin{figure}[htb]
   \epsfysize=5.5cm
   \centerline{\epsffile{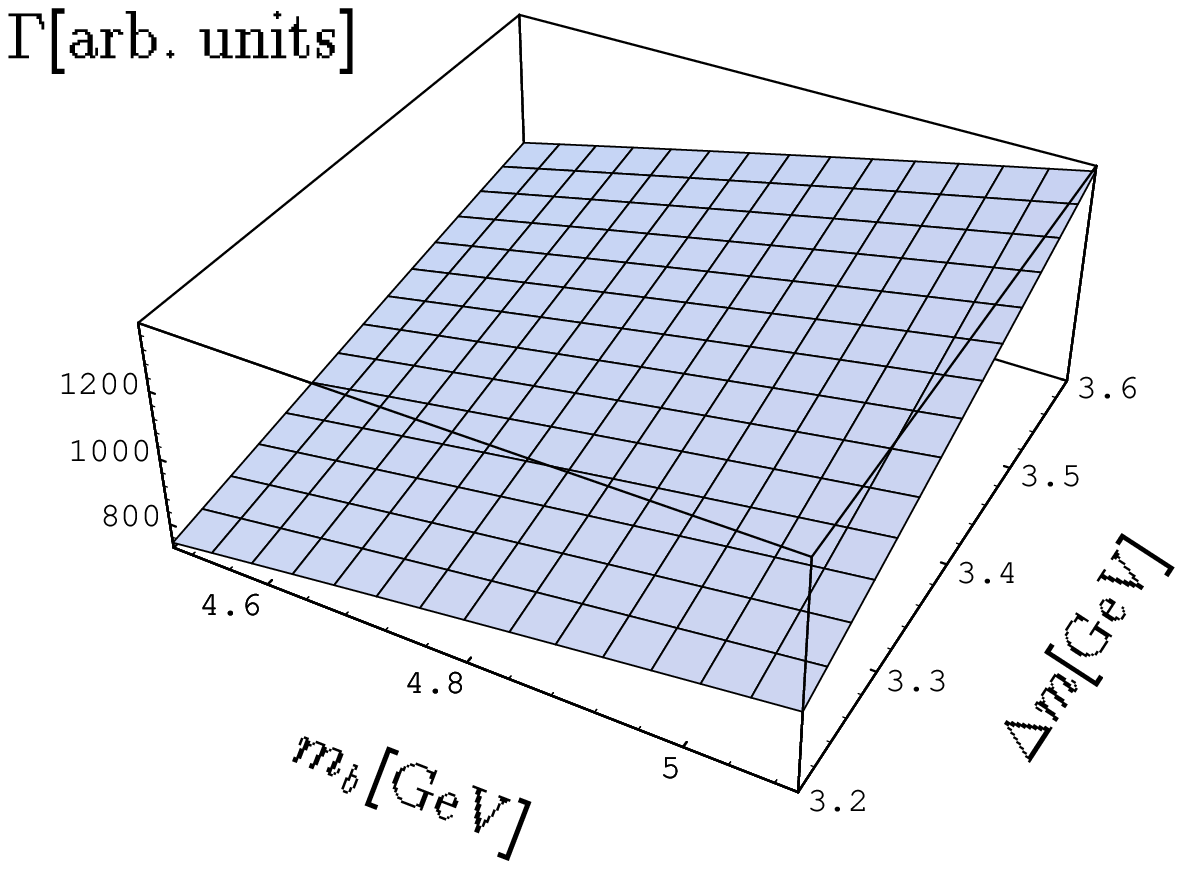}}
\fcaption{Dependence of the inclusive semileptonic decay rate on
$m_b$ and $\Delta m=m_b-m_c$.}
\label{fig:masses}
\end{figure}

Although $\Gamma\sim m_b^5\,f(m_c/m_b)$, the dependence on $m_b$
becomes milder if one chooses $m_b$ and $\Delta m=m_b-m_c$ as
independent variables. This is apparent from Fig.~\ref{fig:masses},
which shows that $\Gamma\sim m_b^{2.3}\,\Delta m^{2.7}$. Moreover,
these variables have essentially uncorrelated theoretical
uncertainties. Whereas $m_b=m_B-\bar\Lambda+\dots$ is mainly
determined by the $\bar\Lambda$ parameter of the HQET\cite{FNL}, the
mass difference $\Delta m$ obeys the expansion\cite{FaNe}
\begin{equation}
   \Delta m = (\bar m_B - \bar m_D)\,\bigg\{ 1
   + {(-\lambda_1)\over 2\bar m_B\bar m_D} + \dots \bigg\}
   = (3.40\pm 0.03\pm 0.03)~\mbox{GeV} \,,
\end{equation}
i.e.\ it is sensitive to the kinetic energy parameter $\lambda_1$.
Here $\bar m_B=5.31$~GeV and $\bar m_D=1.97$~GeV denote the
``spin-averaged'' meson masses, e.g.\ $\bar m_B=\frac{1}{4}(m_B+3
m_{B^*})$.

I think that theoretical uncertainties of 60~MeV on $\Delta m$ and
200~MeV on $m_b$ are reasonable; values much smaller than this are
probably too optimistic. This leads to
\begin{equation}
   \bigg( {\delta\Gamma\over\Gamma} \bigg)_{\rm masses}
   = \sqrt{ \bigg( 0.10\,{\delta m_b\over 200~\mbox{MeV}} \bigg)^2
   + \bigg( 0.05\,{\delta\Delta m\over 60~\mbox{MeV}} \bigg)^2 }
   \simeq 11\% \,.
\end{equation}

\subsubsection{Perturbative corrections}

This is the most subtle part of the analysis. The semileptonic rate
is known exactly only to order\cite{fgrefs} $\alpha_s$, though a
partial calculation of the coefficient of order $\alpha_s^2$
exists\cite{LSW}. The result is
\begin{equation}
   {\Gamma\over\Gamma_{\rm tree}} = 1 - 1.67\,{\alpha_s(m_b)\over\pi}
   - (1.68\beta_0+\dots)\,\bigg( {\alpha_s(m_b)\over\pi} \bigg)^2
   + \dots \,.
\end{equation}
The one-loop correction is moderate; it amounts to about $-11\%$. Of
the two-loop coefficient, only the part proportional to the
$\beta$-function coefficient $\beta_0=11-\frac{2}{3} n_f$ is known.
For $n_f=3$ light quark flavours, this term is $1.68\beta_0\simeq
15.1$ and gives rise to a rather large correction of about $-6\%$.
One may take this as an estimate of the perturbative uncertainty. The
dependence of the result on the choice of the renormalization scale
and scheme has been investigated, too, and found to be of
order\cite{BaNi} 6\%.

Yet, the actual perturbative uncertainty may be larger than that. A
subset of higher-order corrections, the so-called renormalon
contributions of the form $\beta_0^{n-1}\alpha_s^n$, can be summed to
all orders in perturbation theory, leading to\cite{BaBB}
$\Gamma/\Gamma_{\rm tree}=0.77\pm 0.05$, which is equivalent to
choosing the rather low scale $M\simeq 1$~GeV in (\ref{Gamsl}). This
estimate is 12\% lower than the one-loop result.

These considerations show that there are substantial perturbative
uncertainties in the calculation of the semileptonic decay rate. They
could only be reduced with a complete two-loop calculation, which is
however quite a formidable task. At present, I consider
$(\delta\Gamma/\Gamma)_{\rm pert}\simeq 10\%$ a reasonable estimate.

\subsubsection{Result for $|\,V_{cb}|$}

Adding, as previously, the theoretical errors linearly and taking the
square root, I find
\begin{equation}
   {\delta|\,V_{cb}|\over|\,V_{cb}|} \simeq 10\%
\end{equation}
for the theoretical uncertainty in the determination of $|\,V_{cb}|$
from inclusive decays, keeping in mind that in addition this method
relies on the assumption of global duality. Taking the result of Ball
{\it et al}.\cite{BaBB} for the central value, I quote
\begin{equation}
   |\,V_{cb}| = (0.0398\pm 0.0040)\,\bigg( {B_{\rm SL}\over 10.77\%}
   \bigg)^{1/2}\,\bigg( {\tau_B\over 1.6~\mbox{ps}}
   \bigg)^{-1/2} \,.
\end{equation}
Using the new world averages for the semileptonic branching
ratio\cite{Tomasz}, $B_{\rm SL}=(10.77\pm 0.43)\%$, and for the
average $B$ meson lifetime\cite{Joe}, $\tau_B=(1.60\pm 0.03)$~ps, I
obtain
\begin{equation}
   |\,V_{cb}| = (39.8\pm 0.9_{\rm exp}\pm 4.0_{\rm th})
   \times 10^{-3} \,,
\end{equation}
which is is excellent agreement with the measurement in exclusive
decays reported in (\ref{Vcbexc}). This agreement is gratifying given
the differences of the methods used, and it provides an indirect test
of global quark--hadron duality. Combining the two measurements gives
the final result
\begin{equation}
   |\,V_{cb}| = 0.039\pm 0.002 \,.
\end{equation}

\subsection{Semileptonic branching ratio and charm counting}

The semileptonic branching ratio of the $B$ meson is defined as
\begin{equation}
   B_{\rm SL} = {\Gamma(B\to X\,e\,\bar\nu)\over
   \sum_\ell \Gamma(B\to X\,\ell\,\bar\nu) + \Gamma_{\rm NL}
   + \Gamma_{\rm rare}} \,,
\end{equation}
where $\Gamma_{\rm NL}$ and $\Gamma_{\rm rare}$ are the inclusive
rates for nonleptonic and rare decays, respectively, the latter being
negligible. The main difficulty in calculating $B_{\rm SL}$ is not in
the semileptonic width, but in the nonleptonic one. As mentioned
above, the calculation of nonleptonic decays in the $1/m_Q$ expansion
relies on the strong assumption of local quark--hadron duality.

Measurements of the semileptonic branching ratio have been performed
in various experiments, using both model-dependent and
model-independent analyses. The situation has been summarized by
T.~Skwarnicki\cite{Tomasz} at this Conference. The new world average
is
\begin{equation}
   B_{\rm SL} = (10.77\pm 0.43)\% \,.
\end{equation}
An important aspect in understanding this result is charm counting,
i.e.\ the measurement of the average number $n_c$ of charm hadrons
produced per $B$ decay. The CLEO Collaboration has presented a new
result for $n_c$, which is\cite{Tomasz,ncnew}
\begin{equation}
   n_c = 1.16\pm 0.05 \,.
\end{equation}

In the naive parton model, one finds\cite{Alta} $B_{\rm SL}\sim
15$--16\% and $n_c\simeq 1.15$--1.20. Whereas $n_c$ is in agreement
with experiment, the semileptonic branching ratio is predicted too
large. With the establishment of the $1/m_Q$ expansion, the
nonperturbative corrections to the parton model could be computed and
turned out too small to improve the prediction. This lead Bigi {\it
et al}.\ to conclude that values $B_{\rm SL}<12.5\%$ cannot be
accommodated by theory, thus giving rise to a puzzle referred to as
the ``baffling semileptonic branching ratio''\cite{baff}. The
situation has changed recently, however. Bagan {\it et al}.\ found
indications that higher-order perturbative corrections lower the
value of $B_{\rm SL}$ significantly\cite{BSLnew1}. The exact
order-$\alpha_s$ corrections to the nonleptonic width have been
computed for $m_c\ne 0$, and an analysis of the renormalization scale
and scheme dependence has been performed. In particular, it turns out
that radiative corrections increase the partial width $\Gamma(b\to
c\bar c s)$. This has two effects: it lowers the semileptonic
branching ratio, but at the price of a higher value of $n_c$. The
results in two popular renormalization schemes are\cite{BSLnew2}
\begin{eqnarray}
   B_{\rm SL} &=& \cases{
    12.1\pm 0.7_{-1.2}^{+0.9} \% ;& on-shell scheme, \cr
    11.7\pm 0.7_{-1.3}^{+0.9} \% ;& $\overline{\mbox{\sc ms}}$
    scheme, \cr} \nonumber\\
   && \\
   n_c &=& 1.21\mp 0.04\mp 0.01; ~~\mbox{both schemes}.
    \nonumber
\end{eqnarray}
The errors in the two quantities are anti-correlated. The first error
reflects the uncertainties in the input parameters, whereas the
second one shows the dependence on the renormalization scale, which
is varied in the range $m_b/2<\mu<2 m_b$. Lowering $\mu$ decreases
the value of $B_{\rm SL}$ and vice versa. Note that using a low
renormalization scale is not unnatural; Luke {\it et al}.\ have
estimated that $\mu\simeq 0.3 m_b$ is an appropriate scale in this
case\cite{LSW}. Values $B_{\rm SL}<12\%$ can thus easily be
accommodated. Only a complete order-$\alpha_s^2$ calculation could
reduce the perturbative uncertainties.

\begin{figure}[htb]
   \epsfxsize=6cm
   \centerline{\epsffile{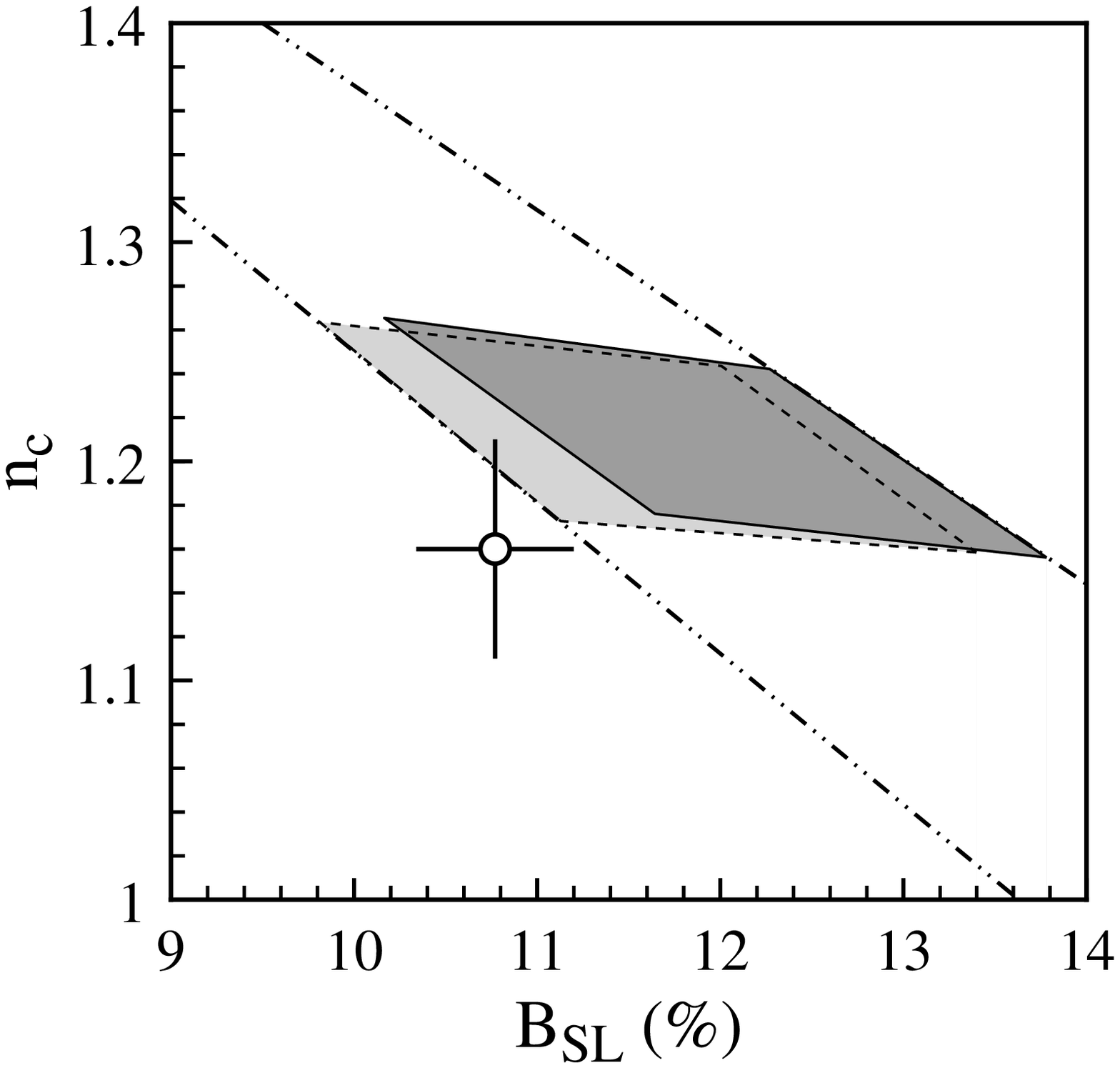}}
\fcaption{Correlation between the semileptonic branching ratio and
charm counting. The dark area is the theoretically allowed region in
the on-shell scheme, whereas the light area refers to the
$\overline{\mbox{\sc ms}}$ scheme\protect\cite{BSLnew2}. The
dash-dotted lines indicate the allowed region if the calculation of
$\Gamma(b\to c\bar c s)$ is ignored and this partial rate is treated
as a free parameter\protect\cite{Buch}. The data point shows the
world average for $B_{\rm SL}$ and the new CLEO result for $n_c$
presented at this Conference.}
\label{fig:BSL}
\end{figure}

The above discussion shows that it is the combination of a low
semileptonic branching ratio and a low value of $n_c$ that
constitutes a potential problem. This is illustrated in
Fig.~\ref{fig:BSL}, which is an updated version of a figure shown in
a recent work of Buchalla {\it et al}.\cite{Buch}. With the new
experimental and theoretical results for $B_{\rm SL}$ and $n_c$, only
a small discrepancy remains between theory and experiment. It has
been argued that the current experimental value of $n_c$ may depend
on model assumptions about the production of charm hadrons, which are
sometimes questionable\cite{Buch,FDW}. Another possibility, which has
been pointed out by Palmer and Stech\cite{PaSt} and
others\cite{FaWi}$^-$\cite{BGM}, is that local quark--hadron duality
could be violated in nonleptonic $B$ decays. If so, this will most
likely happen in the channel $b\to c\bar c s$, where the energy
release, $E=m_B - m_{X(c\bar cs)}$, is of order 1.5~GeV or smaller.
If the discrepancy between theory and experiment persists, this
possibility should be taken seriously before a ``new physics''
explanation\cite{Kagan,Rosz} is advocated.

For completeness, I briefly discuss the semileptonic branching ratio
for $B$ decays into a $\tau$ lepton, which is suppressed by phase
space. The ratio of the semileptonic rates for decays into $\tau$
leptons and into electrons can be calculated reliably. The result
is\cite{FLNN}
\begin{equation}
   B(B\to X\,\tau\,\bar\nu_\tau) = (2.32\pm 0.23)\%\,
   \times {B_{\rm SL}\over 10.77\%} = (2.32\pm 0.25)\% \,.
\end{equation}
It is in good agreement with the new world average\cite{Tomasz}
\begin{equation}
   B(B\to X\,\tau\,\bar\nu_\tau) = (2.60\pm 0.32)\% \,.
\end{equation}

\subsection{Lifetime ratios of $b$-hadrons}

The $1/m_Q$ expansion predicts that the lifetimes of all
$b$-flavoured hadrons agree up to nonperturbative corrections
suppressed by at least two powers of $1/m_b$. This prediction can be
tested with new high precision data, which have been summarized by
J.~Kroll\cite{Joe} at this Conference.

\subsubsection{Lifetime ratio for $B^-$ and $B^0$}

\begin{figure}[htb]
   \epsfysize=3.0cm
   \centerline{\epsffile{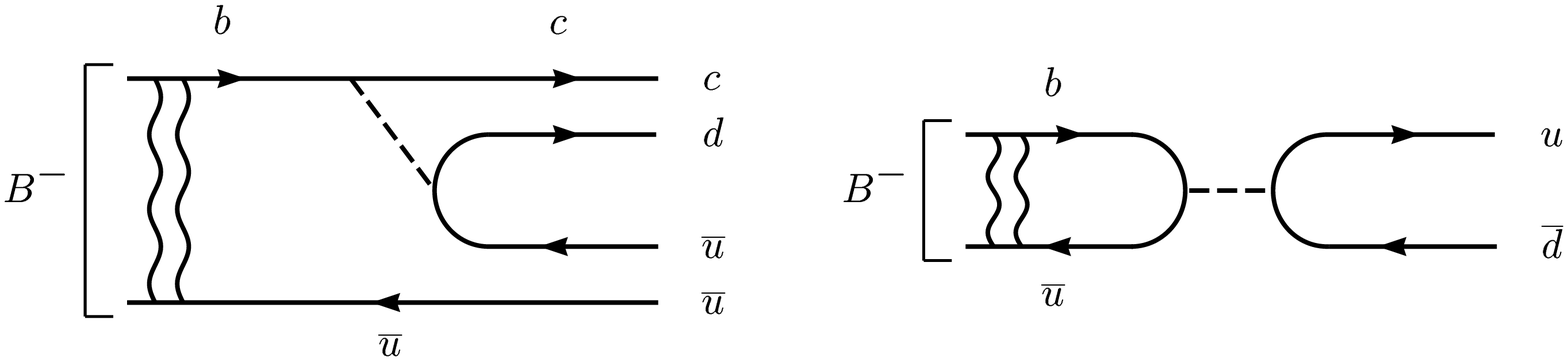}}
\fcaption{Interference and weak-annihilation contributions to the
lifetime of the $B^-$ meson. The interference effect in the first
diagram arises from the presence of two identical $\bar u$ quarks in
the final state.}
\label{fig:WAint}
\end{figure}

The lifetimes of the charged and neutral $B$ mesons differ at order
$1/m_b^3$ in the heavy quark expansion. The corresponding corrections
arise from effects referred to as interference and weak
annihilation\cite{Gube,ShiV}. They are illustrated in
Fig.~\ref{fig:WAint}. In the operator language, these spectator
effects are represented by hadronic matrix elements of local
four-quark operators of the type
\begin{equation}
   \langle\bar b\Gamma q\bar q\Gamma b\rangle_B
   \sim f_B^2\,m_B \sim \Lambda^3 \,,
\end{equation}
where the vacuum insertion approximation\cite{SVZ} has been used. It
turns out that interference gives rise to the dominant corrections
(weak annihilation is strongly CKM suppressed), which decrease the
decay rate for $B^-$, i.e.\ enhance its lifetime. The result
is\cite{liferef}
\begin{equation}\label{DeltaGam}
   \Delta\Gamma_{\rm int}(B^-) = {G_F^2 m_b^5\over 192\pi^3}\,
   |\,V_{cb}|^2\,16\pi^2\,{f_B^2\,m_B\over m_b^3}\,\zeta_{\rm QCD}
   \,,
\end{equation}
where
\begin{equation}
   \zeta_{\rm QCD} = 2 c_+^2(m_b) - c_-^2(m_b)
   = \cases{ \phantom{+} 1 ;& at tree level, \cr
             -0.6 ;& with QCD corrections. \cr}
\end{equation}
After including short-distance corrections to the four-fermion
interactions the interference becomes destructive. The numerical
result is
\begin{equation}
   {\tau(B^-)\over\tau(B^0)} \simeq 1 + 0.04\,
   \bigg( {f_B\over 180~\mbox{MeV}} \bigg)^2 \,,
\end{equation}
consistent with the experimental value\cite{Joe}
\begin{equation}
   {\tau(B^-)\over\tau(B^0)} = 1.02\pm 0.04 \,.
\end{equation}

The theoretical prediction depends on the vacuum saturation
assumption\cite{SVZ}, which has been criticized. Chernyak has
estimated that nonfactorizable contributions can be as large as 50\%
of the factorizable ones\cite{Chern}. Another important observation
is the following one: in nonleptonic decays, spectator effects
appearing at order $1/m_b^3$ are enhanced by a factor $16\pi^2$
resulting from the two-body versus three-body phase space. In fact,
the scale of the correction in (\ref{DeltaGam}) is unexpectedly
large:
\begin{equation}
   16\pi^2\,{f_B^2\,m_B\over m_b^3}\simeq
   \bigg( {4\pi f_B\over m_b} \bigg)^2 \simeq 0.2 \,.
\end{equation}
The presence of this phase-space enhancement factor leads to a
peculiar structure of the $1/m_Q$ expansion for nonleptonic rates,
which can be displayed as follows:
\begin{equation}
   \Gamma \sim \Gamma_0\,\Bigg\{ 1 + \bigg( {\Lambda\over m_Q}
   \bigg)^2 + \bigg( {\Lambda\over m_Q} \bigg)^3 + \dots
   + 16\pi^2\,\bigg[ \bigg( {\Lambda\over m_Q} \bigg)^3
   + \bigg( {\Lambda\over m_Q} \bigg)^4 + \dots \bigg] \Bigg\} \,.
\end{equation}
Numerically, the terms of order $16\pi^2\,(\Lambda/m_Q)^3$ are more
important than the ones of order $(\Lambda/m_Q)^2$. I draw two
conclusion from this observation: it is important to include this
type of $1/m_b^3$ corrections to all predictions for nonleptonic
rates; there is a challenge to calculate the hadronic matrix elements
of four-quark operators with high accuracy. Lattice calculations
could help to improve the existing estimates of such matrix elements.

\subsubsection{Lifetime ratio for $B_s$ and $B_d$}

The lifetimes of the two neutral mesons $B_s$ and $B_d$ differ by
corrections that are due to spectator effects referred to as $W$
exchange. They are smaller than the interference effects discussed
above. The theoretical prediction is\cite{liferef}
\begin{equation}
   {\tau(B_s)\over\tau(B_d)} = 1\pm O(1\%) \,,
\end{equation}
consistent with the experimental value\cite{Joe}
\begin{equation}
   {\tau(B_s)\over\tau(B_d)} = 1.01\pm 0.07 \,.
\end{equation}
Note that $\tau(B_s)$ denotes the average lifetime of the two $B_s$
states.

\subsubsection{Lifetime ratio for $\Lambda_b$ and $B^0$}

Although differences between the lifetimes of heavy mesons and
baryons start at order $1/m_b^2$, the main effects come again at
order $1/m_b^3$. However, here one encounters the problem that the
matrix elements of four-quark operators are needed not only between
meson states (where the vacuum saturation approximation may be used),
but also between baryon states. Very little is known about such
matrix elements. Bigi {\it et al}.\ have adopted a simple
nonrelativistic quark model and conclude that\cite{liferef}
\begin{equation}
   {\tau(\Lambda_b)\over\tau(B^0)} = 0.90\mbox{--}0.95 \,.
\end{equation}
The experimental result for this ratio is significantly
lower\cite{Joe}:
\begin{equation}
   {\tau(\Lambda_b)\over\tau(B^0)} = 0.76\pm 0.05 \,.
\end{equation}
In my opinion, a dedicated theoretical effort to understand this
result is desirable. In view of the above discussion, one should
first question (and improve) the calculation of baryonic matrix
elements of four-quark operators, then question the vacuum saturation
approximation, and finally question the validity of local
quark--hadron duality.

\section{Determinations of $\alpha_s$ from $\Upsilon$ spectroscopy}

Before summarizing, I want to touch upon a topic not related to $B$
decays. The large mass of the $b$ quark makes it possible to describe
the spectrum and properties of $(b\bar b)$ bound states with high
accuracy, using the heavy-quark expansion. From a comparison with
experiment, it is then possible to extract the strong coupling
constant $\alpha_s$. Analyses of this type have been performed based
on lattice calculations and QCD sum rules. The current status of the
determination of $\alpha_s$ from calculations of the $\Upsilon$
spectrum in lattice gauge theory has been summarized by
C.~Michael\cite{Michael} at this Conference. When translated into a
value of $\alpha_s(m_Z)$ in the $\overline{\mbox{\sc ms}}$ scheme, a
conservative result is\cite{CMich,Weisz}
\begin{equation}
   \alpha_s(m_Z) = 0.112\pm 0.007 \,.
\end{equation}
A more precise value, $\alpha_s(m_Z) = 0.115\pm 0.002$, has been
reported by the NRQCD Collaboration\cite{Davies}; the small error has
been criticized, however\cite{CMich,Weisz}.

Voloshin has performed an analysis of the $\Upsilon$ spectrum using
QCD sum rules, including a resummation of large Coulomb corrections
to all orders in perturbation theory\cite{Voloas}. He quotes
$\alpha_s(1~\mbox{GeV})=0.336\pm 0.011$, which translates into
\begin{equation}
   \alpha_s(m_Z) = 0.109\pm 0.001 \,.
\end{equation}
The very small error may have been underestimated. It is important to
understand better the sources of theoretical uncertainty before this
result can be trusted.

Despite such reservations, it looks promising that ultimately the
$\Upsilon$ system may provide one of the best ways to measure
$\alpha_s$ with high precision at low energies.

\section{Summary and conclusions}

I have reviewed the status of the theory of weak decays of heavy
flavours, concentrating on topics relevant to current experiments.
Weak decays play a unique role in testing the Standard Model at low
energies. Ultimately, a precise determination of the parameters of
the flavour sector (elements of the Cabibbo--Kobayashi--Maskawa
matrix and quark masses) will help to explore such intricate
phenomena as CP violation, and is crucial in searches for new physics
beyond the Standard Model.

Exclusive semileptonic decays mediated by the heavy-quark transition
$b\to c\,\ell\,\bar\nu$ are of particular importance, as they allow
for a model-independent description provided by heavy-quark symmetry
and the heavy-quark effective theory. These concepts can now be
tested with detailed measurements of the form factors in the decay
$B\to D^*\ell\,\bar\nu$. The most striking result of the analysis of
this decay is a very precise determination of the strength of $b\to
c$ transitions: $|\,V_{cb}|=(38.6\pm 2.5)\times 10^{-3}$. In the
future, studies of the related decays $B\to D\,\ell\,\bar\nu$,
$B_s\to D_s^{(*)}\ell\,\bar\nu$ and
$\Lambda_b\to\Lambda_c\ell\,\bar\nu$ may provide further tests of
heavy-quark symmetry and teach us about the dependence of the
Isgur--Wise function on the flavour and spin quantum numbers of the
light degrees of freedom in a heavy hadron.

The discovery of exclusive semileptonic decays into charmless final
states is a milestone on the way towards a precise determination of
$|\,V_{ub}|$. Currently, new theoretical ideas are being discussed,
and existing approaches are being refined, which may help to get a
better handle on the calculation of heavy-to-light transition form
factors. These form factors also appear in the description of rare
radiative decays, which are mediated by flavour-changing neutral
currents and in the Standard Model proceed through loop (penguin)
diagrams.

The second part of my talk was devoted to inclusive decays of
$b$-flavoured hadrons. The theoretical description of inclusive rates
is based on the $1/m_Q$ expansion, which is a ``Minkowskian'' version
of the operator product expansion. It can be derived from QCD if one
accepts the hypothesis of quark--hadron duality. Duality is an
important concept in QCD phenomenology, which however cannot be
derived yet from first principles, so it needs to be tested. The
measurement of the inclusive semileptonic decay rate of the $B$ meson
provides an alternative way to determine $|\,V_{cb}|$, which leads to
$|\,V_{cb}|=(39.8\pm 4.1)\times 10^{-3}$, in excellent agreement with
the value obtained from exclusive decays. This agreement provides an
indirect test of global quark--hadron duality. The theoretical
calculation of the semileptonic branching ratio, $B_{\rm SL}$,
suffers from a sizable renormalization-scheme dependence, which does
not allow to state a discrepancy between the data and theory for
$B_{\rm SL}$ alone. However, the combination of a low value of
$B_{\rm SL}$ and a low value of $n_c$, the number of charm hadrons
per $B$ decay, remains a problem that deserves further investigation.

A particularly clean test of the heavy-quark expansion is provided by
the study of lifetime ratios of $b$-flavoured hadrons. This tests the
assumption of local quark--hadron duality, as well as our capability
to evaluate the hadronic matrix elements of the operators appearing
in the $1/m_b$ expansion. For the lifetime ratios
$\tau(B_d)/\tau(B_s)$ and $\tau(B^-)/\tau(B^0)$ there is good
agreement between theory and experiment, although the data have not
yet reached the precision required to perform a stringent test of the
theoretical predictions. However, in the case of the lifetime ratio
$\tau(\Lambda_b)/\tau(B^0)$ there are indications for a significant
discrepancy; the data indicate much larger power corrections than
anticipated by theory. A better understanding of this discrepancy, if
it persists, is most desirable.

Given the limitations in space and time, the material covered in this
talk is only a selection of current topics in heavy flavour physics.
With continuous advances on the theoretical front, and with ongoing
experimental efforts and the construction of new facilities ($B$
factories), this field will remain of great interest and will
continue to provide us with new exciting results in the future.

\vspace{0.5cm}
{\it Acknowledgements:\/}
I am grateful to Patricia Ball, Guido Martinelli, Chris Sach\-raj\-da
and Daniel Wyler for helpful discussions and suggestions. I would
also like to thank my experimental colleagues Tomasz Skwarnicki and
Joe Kroll for numerous exchanges before, during and after this
Conference.


\begin{thebibliography}{999}
\bibitem {RKmodels}
A compilation of quark-model results on $R_{K^*}$ can be found in
Ref.\cite{CLRK}.
\end{thebibliography}
\end{document}